%% file: main.tex
\documentclass{Interspeech}

\interspeechcameraready

\title{xLSTM-SENet: xLSTM for Single-Channel Speech Enhancement}

\author[affiliation={1}]{Nikolai Lund}{Kühne}
\author[affiliation={1}]{Jan}{Østergaard}
\author[affiliation={1,2}]{Jesper}{Jensen}
\author[affiliation={1}]{Zheng-Hua}{Tan}

\affiliation{Department of Electronic Systems}{Aalborg University}{Denmark}
\affiliation{Oticon A/S}{Copenhagen}{Denmark}
\email{\{nlk,jo,jje,zt\}@es.aau.dk}
\keywords{xLSTM, speech enhancement, LSTM, Mamba}

\usepackage{comment}

\input{Preamble}

\begin{document}

\maketitle

\input{Text/00Abstract}
\input{Text/01Introduction}
\input{Text/02Method}

\input{Text/03Experiments}

\input{Text/04Results}
\input{Text/05Conclusion}

% COMMENT IN TO GET BIB
\bibliographystyle{IEEEtran}
\bibliography{mybib}

\end{document}

%% file: Preamble.tex
\usepackage{balance}                % Balances the columns of the last page
\usepackage{dirtytalk} %benytte \say til at lave pæne anførselstegn
\usepackage{siunitx}						% Flot og konsistent praesentation af tal og enheder med \si{enhed} og \SI{tal}{enhed}
\usepackage[official]{eurosym} 

\usepackage{soul}
\usepackage{makecell}

\makeatletter
\newcommand*{\bigs}[1]{{\hbox{$\left#1\vbox to10\p@{}\right.\n@space$}}}
\makeatother

\makeatletter
\newcommand*{\biggs}[1]{{\hbox{$\left#1\vbox to17\p@{}\right.\n@space$}}}
\makeatother

\usepackage[english]{babel}
\usepackage{blindtext}

\usepackage{tikz}  %Pakke til plot af funktioner og grafer
\usetikzlibrary{decorations.pathreplacing,calligraphy}
\usepackage[framemethod=tikz]{mdframed}
\usetikzlibrary{shapes.geometric, arrows, arrows.meta}
\usepackage{graphicx}
\usepackage{float}
\usepackage{booktabs}
\usepackage{multirow}
\usepackage{dblfloatfix}
\DeclareSIUnit[quantity-product = ]\percent{\char`\%}
\usepackage{comment}
\usetikzlibrary{decorations.text}
\usetikzlibrary{shapes.geometric, arrows}
\usetikzlibrary{positioning}
\usepackage{adjustbox}

\usetikzlibrary{intersections}
\def\,{\mskip\thinmuskip} \def\!{\mskip-\thinmuskip}
\def\,{\mskip\thickmuskip} \def\;{\mskip+\thickmuskip}

\usepackage{cite}
\usepackage{amsmath,amssymb,amsfonts,mathtools}
\usepackage{algorithmic}
\usepackage{graphicx}
\usepackage{textcomp}
\usepackage{xcolor}
\def\BibTeX{{\rm B\kern-.05em{\sc i\kern-.025em b}\kern-.08em
    T\kern-.1667em\lower.7ex\hbox{E}\kern-.125emX}}

\usepackage{bm}
\usepackage{comment}

\usepackage{pifont} % Checkmark
\makeatletter
\newcommand*\bigcdot{\mathpalette\bigcdot@{.5}}
\newcommand*\bigcdot@[2]{\mathbin{\vcenter{\hbox{\scalebox{#2}{$\m@th#1\bullet$}}}}}
\makeatother
\usepackage{tikz}  %Pakke til plot af funktioner og grafer
\usepackage{url}

\definecolor{mygreen}{RGB}{127,201,127}
\definecolor{mypurple}{RGB}{190,174,212}
\definecolor{myorange}{RGB}{253,192,134}

%% file: Text/00Abstract.tex
\begin{abstract}
While attention-based architectures, such as Conformers, excel in speech enhancement, they face challenges such as scalability with respect to input sequence length. In contrast, the recently proposed Extended Long Short-Term Memory (xLSTM) architecture offers linear scalability. However, xLSTM-based models remain unexplored for speech enhancement. This paper introduces xLSTM-SENet, the first xLSTM-based single-channel speech enhancement system. A direct comparative analysis reveals that xLSTM—and notably, even LSTM—can match or outperform state-of-the-art Mamba- and Conformer-based systems across various model sizes in speech enhancement on the VoiceBank+Demand dataset. Through ablation studies, we identify key architectural design choices such as exponential gating and bidirectionality contributing to its effectiveness. Our best xLSTM-based model, xLSTM-SENet2, outperforms state-of-the-art Mamba- and Conformer-based systems of similar complexity on the Voicebank+DEMAND dataset.
\end{abstract}

%% file: Text/01Introduction.tex
\section{Introduction}
Real-world speech signals are often disrupted by noise, which degrades performance in hearing assistive devices \cite{kolbaek2016speech}, automatic speech recognition systems \cite{chen15o_interspeech}, and for speaker verification \cite{shon19b_interspeech}. The process of removing background noise and enhancing the quality and intelligibility of the desired speech signal is known as speech enhancement (SE). Given the wide range of applications, SE has garnered significant research interest.

Single-channel SE methods leveraging deep learning \cite{kolbaek2016speech} encompass a selection of architectures, such as recurrent Long Short-Term Memory (LSTM) networks \cite{tesch22_interspeech}, convolutional neural networks (CNNs) \cite{fu2016snr, kolbaek2020loss}, generative adversarial networks (GANs) \cite{michelsanti17_interspeech, fu2019metricgan, metricgan+, cmgan}, and diffusion models \cite{lu2022conditional, richter2023speech,gonzalez2024investigating}. Recently, Conformer-based models have demonstrated impressive SE performance, achieving state-of-the-art results \cite{mp-senet, cmgan} on the VoiceBank+Demand dataset \cite{voicebank, demand}. However, models based on scaled dot-product attention, such as Transformers and Conformers, face challenges with scalability with respect to input sequence length \cite{deoliveira22_interspeech}, and they require a lot of data to train \cite{gong21b_interspeech}.

To address the inherent limitations of attention-based models, Mamba \cite{mamba}, a sequence model integrating the strengths of CNNs, RNNs, and state space models, has recently emerged. Mamba has demonstrated competitive or superior performance relative to Transformers across various tasks, including audio classification \cite{audiomamba} and SE \cite{semamba}. 

On the other hand, recurrent neural networks (RNNs), particularly LSTMs \cite{lstm}, also offer several advantages over attention-based models, including: (i) linear scalability instead of quadratic, in terms of computational complexity with respect to the input sequence length, and (ii) reduced runtime memory requirements, as they do not require storage of the full key-value (KV) cache. In contrast, attention-based models such as Transformers and Conformers necessitate significant memory overhead for KV storage.
Despite these advantages, LSTMs suffer from critical drawbacks: (i) inability to revise storage decisions, (ii) reliance on scalar cell states, constraining storage capacity, and (iii) memory mixing preventing parallelizability \cite{xlstm}. Consequently, LSTMs have been less utilized in recent deep learning-based SE systems, with only a few exceptions \cite{tesch22_interspeech, tesch2023multi}.

\begin{figure*}[thb]
    \centering
    \input{Pix/SExLSTM}
    \caption{Overall structure of our proposed xLSTM-SENet with parallel magnitude and phase spectra denoising.}
    \label{fig: SExLSTM}
\end{figure*}
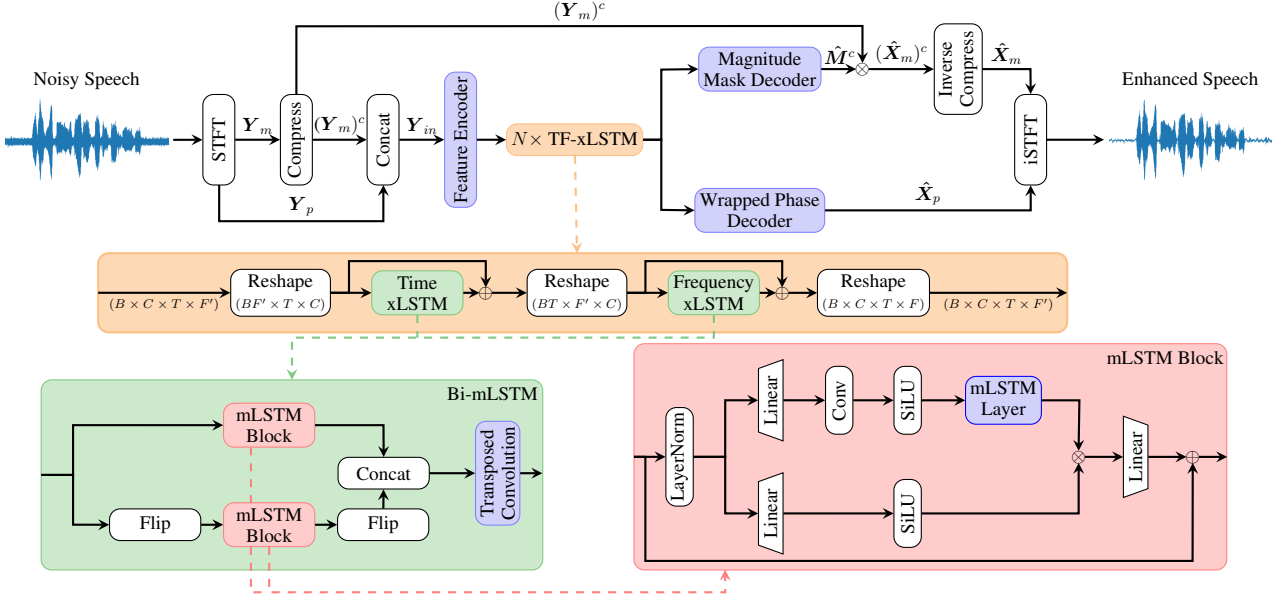
Recently, the Extended Long Short-Term Memory architecture (xLSTM) \cite{xlstm} was proposed to overcome the limitations of LSTMs. By incorporating exponential gating, matrix memory, and improved normalization and stabilization mechanisms while eliminating traditional memory mixing, xLSTM introduces two new fundamental building blocks: sLSTM and mLSTM. The xLSTM architecture has shown competitive performance across tasks such as natural language processing \cite{xlstm}, computer vision \cite{visionlstm}, and audio classification \cite{audioxlstm}. However, while xLSTM adds increased memory via matrix memory and an improved ability to revise storage decisions via exponential gating, the potential advantages of these additions over LSTM have yet to be assessed for SE.

In this work, we propose an xLSTM-based SE system (xLSTM-SENet), which is the first single-channel SE system utilizing xLSTM. The system architecture is illustrated in \autoref{fig: SExLSTM}. 
Systematic comparisons of our proposed xLSTM-SENet with Mamba-, and Conformer-based counterparts across various model sizes, show that xLSTM-SENet matches the performance of state-of-the-art Mamba- and Conformer-based systems on the VoiceBank+Demand dataset \cite{voicebank, demand}. Intriguingly, upon an in depth investigation, we find that LSTMs can match or even outperform xLSTM, Mamba, and Conformers on the VoiceBank+Demand dataset.
Additionally, we perform detailed ablation studies to explore the importance of multiple architectural design choices, quantifying their impact on overall performance. 
Finally, our best configured xLSTM-based model, xLSTM-SENet2, outperforms state-of-the-art Mamba- and Conformer-based systems of similar complexity on the Voicebank+DEMAND dataset. Code is publicly available.\footnote{\url{https://github.com/NikolaiKyhne/xLSTM-SENet}}

%% file: Pix/SExLSTM.tex
\usetikzlibrary{shapes.geometric, arrows, positioning, calc}
\tikzstyle{block} = [rectangle, rounded corners, minimum width=2cm, minimum height=0.7cm,text centered, draw=black]
\tikzstyle{arrow} = [thick,->,>=stealth]
\usetikzlibrary{backgrounds,scopes}   %<------- Load libraries

\begin{tikzpicture}[auto, node distance=1.5cm,>=latex', scale=0.6, every node/.style={transform shape}, font=\large]
% Input waveform
\node (input) {\includegraphics[scale=0.15]{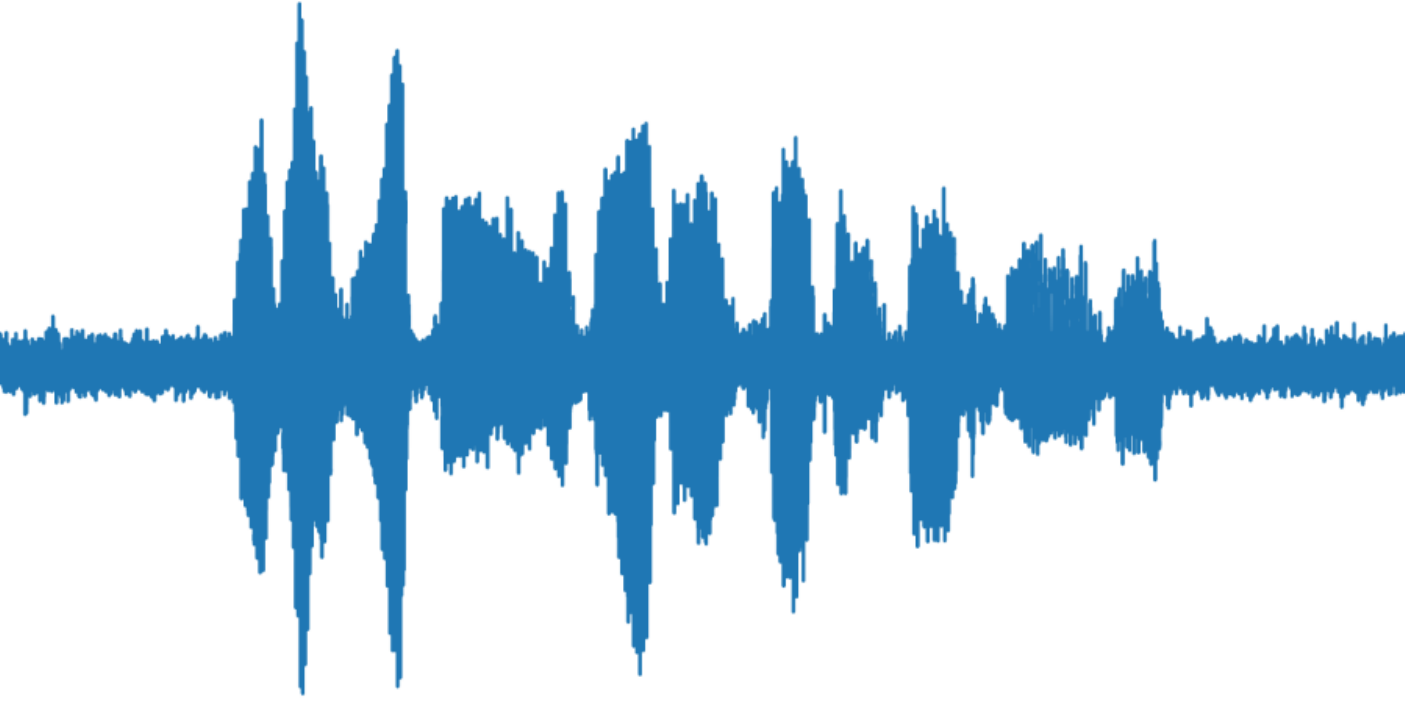}};

\node [above, align=center] at (input.north) {\large Noisy Speech};

% STFT
\node (stft) [block, right=1cm of input, rotate=90, xshift=-1cm] {STFT};
% Compression
\node (compress) [block, right of=stft, xshift=0.2cm, rotate=90] {Compress};

% Concat
\node (concat) [block, right of=compress, xshift=0.4cm, rotate=90] {Concat};
% Feature Encoder
\node (encoder) [block, right of=concat, xshift=0.2cm, rotate=90, fill=blue!80!gray!20, draw = blue!50] {Feature Encoder};

% TF-xlstm
\node (tfxlstm) [block, right of=encoder, xshift=1cm, fill=myorange!60, draw = myorange] {$N\times$ TF-xLSTM};

% Decoders
% \node [draw, align=center] (NN) at (4, 0) {Alternative Sequence Modelling \\ Neural Architectures};

\node (magdecoder) [align=center, block, above right of=tfxlstm, xshift=3cm, yshift=0.5cm,fill=blue!80!gray!20, draw = blue!50] {Magnitude \\ Mask  Decoder};
\node (phasedecoder) [align=center, block, below right of=tfxlstm, xshift=3cm, yshift=-0.5cm, fill=blue!80!gray!20, draw = blue!50] {Wrapped Phase \\ Decoder};

\node (mult1) [align=center, right of=magdecoder, xshift=0.75cm, yshift=0cm] {$\otimes$};

% Inverse Compression and iSTFT
\node (inversecompress) [align=center, block, right of=mult1, xshift=0.6cm, rotate=90] {Inverse \\ Compress};
\node (istft) [block, right of=tfxlstm, xshift=8.5cm, rotate=90] {iSTFT};

% Output waveform

\node (output) [right of=istft, xshift=2cm] {\includegraphics[scale=0.15]{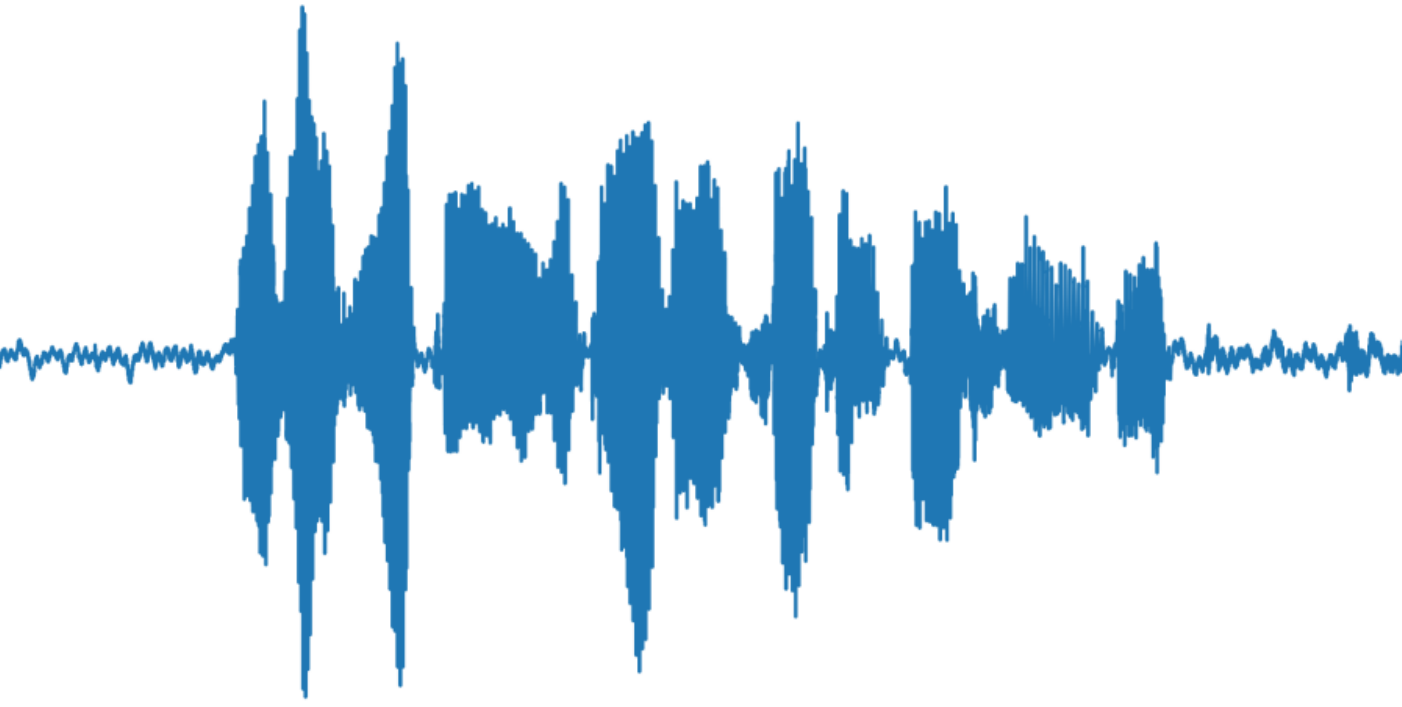}};

\node [above, align=center] at (output.north) {Enhanced Speech};

% Connections
\draw [arrow] (input) -- (stft);

\draw [arrow] (stft) -- node [text width=1.5cm,midway,above,align=center] {$\bm{{Y}}_m$} (compress);

\draw [arrow] (compress) -- node [text width=1.5cm,midway,above,align=center] {($\bm{{Y}}_m)^c$} (concat);

% Add residual connection from STFT to Compress
\draw [arrow] ([yshift=-1cm, xshift=-0.35cm]stft.south) -- ++(0,-0.8) -- node [text width=1.5cm,midway,above,align=center] {$\bm{{Y}}_p$} ++(3.60,0) -- ([yshift=-1cm, xshift=-0.35cm]concat.south);

\draw [arrow] (concat) -- node [text width=1.5cm,midway,above,align=center] {$\bm{{Y}}_{in}$} (encoder);

\draw [arrow] (encoder) -- (tfxlstm);

\draw [arrow] (tfxlstm.east) -- ++(0.40, 0) -- ++(0, 1.55) -- (magdecoder.west);

\draw [arrow] (tfxlstm.east)  -- ++(0.40, 0) -- ++(0, -1.55) -- (phasedecoder.west);

\draw [arrow] (magdecoder) -- node [text width=1.5cm,midway,above,align=center,xshift=0.09cm] {$\bm{\hat{M}}^c$} (mult1.west) -- ++(0.14cm,0);

\draw [arrow] (compress.east) -- ++(0,1.50cm) -- node [text width=1.5cm,midway,above,align=center] {$(\bm{{Y}}_m)^c$} ([yshift=0.7cm]mult1.north) -- ++(0,-0.82cm);

\draw [arrow] ([xshift=-0.14cm] mult1.east)  -- node [text width=1.5cm,midway,above,align=center] {$(\bm{\hat{X}}_m)^c$}(inversecompress);

\draw [arrow] (inversecompress.south) -- node [text width=1.5cm,midway,above,align=center] {$\bm{\hat{X}}_m$} ++(1.05, 0) -- (istft.east);

\draw [arrow] (phasedecoder.east) -- node [text width=1.5cm,midway,above,align=center] {$\bm{\hat{X}}_p$} ++(4.510, 0) -- (istft.west);

\draw [arrow] (istft) -- (output);

\begin{scope}[on background layer]
\draw[rounded corners, thick,  fill=myorange!60, draw = myorange] 
    ($(0.25,-2.5)$) rectangle 
    ($(21.5,-4.25)$);
\end{scope}

\node (reshape1) [align=center,block, fill=white] at (4.25, -3.375) {Reshape\\ \footnotesize$(BF'\times T\times C)$};
\node (timexlstm) [align=center,block, fill=mygreen!40, draw = mygreen, right of=reshape1, xshift=1.5cm] {Time \\ xLSTM};
\node (add1) [align=center, right of=timexlstm] {$\oplus$};
\node (reshape2) [align=center,block, fill=white, right of=timexlstm, xshift=2.cm] {Reshape\\ \footnotesize$(BT\times F'\times C)$};
\node (freqxlstm) [align=center,block, fill=mygreen!40, draw = mygreen, right of=reshape2, xshift=1.5cm] {Frequency \\ xLSTM};
\node (add2) [align=center, right of=freqxlstm] {$\oplus$};
\node (reshape3) [align=center,block, fill=white, right of=freqxlstm, xshift=2.cm] {Reshape\\ \footnotesize$(B\times C\times T \times F)$};

\draw[arrow] (0.25,-3.375) -- node [text width=3.0cm,midway,below,align=center] {\footnotesize$(B\times C\times T\times F')$} (reshape1);
\draw[arrow] (reshape1) -- (timexlstm);
\draw[arrow] (timexlstm) -- (add1.west) -- ++(0.14cm,0);
\draw [arrow] (reshape1.east) -- ++(0.40,0) -- ++(0,0.7cm) -- ++(3.00cm,0) -- ++ ([xshift=-8.7505cm, yshift=2.54cm]add1.north);
\draw[arrow] ([xshift=-0.14cm]add1.east) -- (reshape2);
\draw[arrow] (reshape2) -- (freqxlstm);
\draw[arrow] (freqxlstm) -- (add2.west) -- ++(0.14cm,0);
\draw [arrow] (reshape2.east) -- ++(0.40,0) -- ++(0,0.7cm) -- ++(3.00cm,0) -- ++ ([xshift=-15.255cm, yshift=2.54cm]add2.north);
\draw[arrow] ([xshift=-0.14cm]add2.east) -- (reshape3);
\draw[arrow] (reshape3) -- node [text width=3.0cm,midway,below,align=center] {\footnotesize$(B\times C\times T\times F')$} (21.5,-3.375);
\draw[arrow, dashed, myorange] (tfxlstm) -- ([xshift=0.0cm, yshift=0.35cm]reshape2.north);

\begin{scope}[on background layer]
\draw[rounded corners,fill=mygreen!40, draw=mygreen] 
    ($(-1,-5.3)$) rectangle 
    ($(10,-9.5)$);
\end{scope}

\node (mlstmblock1) [align=center,block, fill=red!20, draw = red!50] at (4, -6.3) {mLSTM\\ Block};
\node (mlstmblock2) [align=center,block, fill=red!20, draw = red!50] at (4, -8.5) {mLSTM\\ Block};
\node (flip1) [align=center,block, fill=white, left of=mlstmblock2, xshift=-1cm] {Flip};
\node (flip2) [align=center,block, fill=white, right of=mlstmblock2, xshift=1cm] {Flip};

\node (concat2) [align=center,block, fill=white, below right of=mlstmblock1, xshift=1.45cm] {Concat};
\node (transpconv) [align=center,block, fill=blue!80!gray!20, draw = blue!50, right of=concat2, rotate=90, yshift=-1.cm] {Transposed\\ Convolution};

\draw[arrow] (-1, -7.4) -- ++ (0.7cm,0) -- ++ (0,1.1cm) -- (mlstmblock1);
\draw[arrow] (-1, -7.4) -- ++ (0.7cm,0) -- ++ (0,-1.1cm) -- (flip1);
\draw[arrow] (flip1) --(mlstmblock2);
\draw[arrow] (mlstmblock2) --(flip2);
\draw[arrow] (flip2) --(concat2);
\draw[arrow] (concat2) --(transpconv);
\draw[arrow] (transpconv) -- ++ (0.98,0);
\draw[arrow] (mlstmblock1.east) -- ++ (1.5,0) -- (concat2.north);
\tikzstyle{block} = [rectangle, rounded corners, minimum width=1.5cm, minimum height=0.6cm,text centered, draw=black]

\begin{scope}[on background layer]
\draw[rounded corners,fill=red!20, draw = red!50] 
    ($(12,-4.5)$) rectangle 
    ($(25,-9.5)$);
\end{scope}
\node (norm) [align=center,block, fill=white, rotate=90] at (13, -7) {LayerNorm};
\node [trapezium, draw, align=center, fill=white, rotate=90, trapezium right angle=70, trapezium left angle=70, minimum width=1.8cm] (lin1)  at (15, -5.75) {Linear};
\node (conv) [align=center,block, fill=white, right of=lin1, rotate=90] {Conv};
\node (silu) [align=center,block, fill=white, right of=conv, rotate=90] {SiLU};
\node (mlstmlayer) [align=center,block, fill=blue!80!gray!20, draw = blue!503,, right of=silu, xshift =0.6cm] {mLSTM \\ Layer};

\node (lin2) [trapezium, draw, align=center, fill=white, rotate=90, trapezium right angle=70, trapezium left angle=70, minimum width=1.8cm] at (15, -8.25) {Linear};
\node (silu2) [align=center,block, fill=white, right of=lin2, rotate=90, yshift=-1.5cm] {SiLU};

\node (mult2) [align=center, right of=norm, xshift=7.25cm, yshift=0cm] {$\otimes$};

\node (lin3) [trapezium, draw, align=center, fill=white, rotate=90, trapezium right angle=110, trapezium left angle=110, minimum width=1.8cm, above of=mult2, yshift=-2.75cm]  {Linear};

\node (add3) [align=center, right of=mult2, xshift=1cm] {$\oplus$};

\draw[arrow] (12,-7) -- (norm);
\draw[arrow] (12,-7) -- ++ (0.25, 0) -- ++ (0, -2.30) -- ++ (12, 0) -- (add3.south) -- ++ (0, 0.12);

\draw[arrow] (norm) -- ++ (1, 0) -- ++ (0, -1.25) -- (lin2);
\draw[arrow] (norm) -- ++ (1, 0) -- ++ (0, 1.25) -- (lin1);
\draw[arrow] (lin1) -- (conv);
\draw[arrow] (conv) -- (silu);
\draw[arrow] (silu) -- (mlstmlayer);
\draw[arrow] (lin2) -- (silu2);
\draw[arrow] (silu2) -- ++ (3.75,0) -- (mult2.south) -- ++ (0, 0.12);
\draw[arrow] (mlstmlayer.east) -- ++ (0.825,0) -- (mult2.north) -- ++ (0, -0.12);
\draw[arrow] ([xshift=-0.14cm]mult2.east) -- (lin3);
\draw[arrow] (lin3) -- (add3.west) -- ++ (0.14, 0);
\draw[arrow] ([xshift=-0.14cm]add3.east) -- ++ (0.62, 0);

%last grey arrows
\draw[arrow, dashed, -, mygreen] (timexlstm.south)  -- ++ (0, -0.50);
\draw[arrow, dashed, mygreen] (freqxlstm.south) -- ++ (0, -0.50) -- ++ (-9.25, 0) -- (4.5, -5.3);
\draw[arrow, dashed, -, red!50] ([xshift=-0.4cm]mlstmblock1.south)  -- ([xshift=-0.4cm]mlstmblock2.north);
\draw[arrow, dashed, red!50] (mlstmblock2.south)  -- ++ (0,-1cm) -- ++ (10cm, 0) -- ++ (0, 0.50cm);
\draw[arrow, dashed, -, red!50] ([xshift=-0.4cm]mlstmblock2.south)  -- ++ (0,-1cm) -- ++ (0.25cm, 0);

%text in scopes
\node[] at (8.9,-5.6) {Bi-mLSTM};
\node[] at (23.65,-4.8) {mLSTM Block};

\end{tikzpicture}

%% file: Text/02Method.tex
\section{Method}

\subsection{Extended long short-term memory}
As mentioned, xLSTM \cite{xlstm} introduces two novel building blocks: sLSTM and mLSTM, to address the limitations of the original LSTM \cite{lstm}. Following Vision-LSTM \cite{visionlstm} and Audio xLSTM \cite{audioxlstm}, we employ mLSTM as the main building block in our SE system. Unlike the sigmoid gating used in traditional LSTMs, mLSTM adopts exponential gating for the input and forget gates, enabling it to better revise storage decisions. Additionally, the scalar memory cell $c \in \mathbb{R}$ is replaced with a matrix memory cell $\bm{C} \in \mathbb{R}^{d \times d}$ to increase storage capacity. Each mLSTM block projects the $D$-dimensional input by an expansion factor $E_f \in \mathbb{N}$ to $d = E_f D$ before projecting it back to $D$-dimensions after being processed by an mLSTM layer.
% At time-step $t$, mLSTM stores a key-value pair $\bm{k}_t,\bm{v}_t\in\mathbb{R}^d$ via \eqref{eq1}, and later the relevant value $\bm{v}_t$ is retrieved by a query $\bm{q}_{t+\tau}\in\mathbb{R}^d$. 
The forward pass of
mLSTM is given by \cite{xlstm}:         
\begin{align}
{\bm{C}_t} \ &= \  {f_t} \ {\bm{C}_{t-1}} \ + \ \label{eq1}
  {i_t} \ {\bm{v}_t \ \bm{k}_t^\top} ,\\
{\bm{n}_t} \ &= \  {f_t} \ {\bm{n}_{t-1}} \ + \ 
  {i_t} \ {\bm{k}_t} ,\\
\bm{h}_t  \ &= \ {\bm{o}_t} \ \odot \ \left({\bm{C}_t} {\bm{q}_t} \ / \ 
\max \left\{ |{\bm{n}_t^\top} {\bm{q}_t}|, 1 \right\} \right)  ,\\
\bm{q}_t \ &= \ \bm{W}_q \ \bm{x}_t \ + \ \bm{b}_q,  & & \\
\bm{k}_t \ &= \ \frac{1}{\sqrt{d}} \bm{W}_k \ \bm{x}_t \ + \ \bm{b}_k,  & &\\
\bm{v}_t \ &= \ \bm{W}_v \ \bm{x}_t \ + \ \bm{b}_v,  & & \\
{i_t} \ &= \ \exp{\left( \bm{w}^\top_{i} \ \bm{x}_t \ + \  b_{i}  \right)} \ ,\\
{f_t} \ &= \ \exp{\left( \bm{w}^\top_{f} \ \bm{x}_t  \ + \ b_{f} \right)} ,\\
\label{eq:mlstm_recurrent_end}
\bm{o}_t \ &= \ \sigma \left( \bm{W_{o}} \ \bm{x}_t \ + \
\bm{b_{o}}\right), \,
\end{align}
where the cell state $\bm{C}_t\in\mathbb{R}^{d\times d}$, $\bm{n}_t,\bm{h}_t\in\mathbb{R}^d$ represent the normalizer state and the hidden state, respectively, $\sigma(\cdot)$ is the sigmoid function, and $\odot$ is element-wise multiplication. The input, forget and output gates are represented by $i_t,f_t\in\mathbb{R}$ and $\bm{o}_t\in\mathbb{R}^d$, respectively, while $\bm{W}_q, \bm{W}_k, \bm{W}_v \in \mathbb{R}^{d\times d}$ are learnable projection matrices and $\bm{b}_q, \bm{b}_k, \bm{b}_v \in \mathbb{R}^{d}$ are the respective biases for the query, key and value vectors. Finally, $\bm{w}_i, \bm{w}_f\in\mathbb{R}^d$, and $\bm{W_o}\in\mathbb{R}^{d\times d}$ represent the weights between the input $\bm{x}_t$ and the input, forget, and output gate, respectively, and ${b}_i, b_f\in \mathbb{R}$ and $\bm{b_o} \in \mathbb{R}^{d}$ are their biases.
Unlike LSTM and sLSTM, there are no interactions between hidden states from one time step to the following in mLSTM (i.e. no memory mixing). This means multiple memory cells and multiple heads are equivalent, allowing the forward pass to be parallelized. 

\subsection{xLSTM-SENet: speech enhancement with xLSTMs}
% Following SEMamba \cite{semamba}, we integrate xLSTM into the MP-SENet architecture \cite{mp-senet} by replacing the Conformers in MP-SENet with xLSTM blocks as shown in \autoref{fig: SExLSTM}. 
For a direct comparison with the state-of-the-art dual-path \cite{luo2020dual} SE systems: SEMamba \cite{semamba} and MP-SENet \cite{mp-senet}, we integrate xLSTM into the MP-SENet architecture by replacing the Conformer blocks with xLSTM blocks as shown in \autoref{fig: SExLSTM}. 
We use the MP-SENet architecture since it facilitates joint denoising of magnitude and phase spectra, and has shown superior performance compared to other time-frequency (TF) domain SE methods \cite{mp-senet}.

\subsubsection{Model structure}
\textbf{Model overview:} As shown in \autoref{fig: SExLSTM}, our proposed xLSTM-SENet architecture follows an encoder-decoder structure. 
Given the noisy speech waveform $\bm{y}\in\mathbb{R}^D$, let $\bm{Y}=\bm{Y}_m \cdot \mathrm{e}^{j\bm{Y}_p}\in\mathbb{C}^{T\times F}$ (where $T$ and $F$ represent time and frequency dimensions, respectively) denote the corresponding complex spectrogram obtained through a short-time Fourier transform (STFT).
We stack the wrapped phase spectrum $\bm{Y}_p\in\mathbb{R}^{T\times F}$ and the compressed magnitude spectrum $(\bm{Y}_m)^c\in\mathbb{R}^{T\times F}$ (extracted by applying power-law compression \cite{powerlawcompression} with compression factor $c=0.3$ as in \cite{mp-senet}) to create an input $\bm{Y}_{in}\in\mathbb{R}^{T\times F\times 2}$ to the feature encoder. The feature encoder encodes the input into a compressed TF-domain representation, which is subsequently fed to a stack of $N$ TF-xLSTM blocks. 
% Then $\bm{Y}_p\in\mathbb{R}^{T\times F}$ is the wrapped phase spectrum, and by applying power-law compression \cite{powerlawcompression} with compression factor $0<c\in\mathbb{R}$, we extract the compressed magnitude spectrum $(\bm{Y}_m)^c\in\mathbb{R}^{T\times F}$.   
% The stacked compressed magnitude spectrum and wrapped phase spectrum $\bm{Y}_{in}=(\bm{Y}_m)^c \oplus \bm{Y}_p\in\mathbb{R}^{T\times F\times 2}$, where $\oplus$ is the concantation operator, is encoded into a compressed TF-domain representation, which is subsequently fed to a stack of $N$ TF-xLSTM blocks. 
Each TF-xLSTM block comprises a time and frequency xLSTM block, capturing temporal and frequency dependencies, respectively. 
Similar to SEMamba \cite{semamba}, we use a bidirectional architecture (Bi-mLSTM) for these blocks. Hence, the output $\bm{\upsilon}$ of the time and frequency xLSTM blocks is:
\begin{align}
    \bm{\upsilon} = \mathrm{Conv1d}(\mathrm{mLSTM}(\bm{\varepsilon}) \oplus \mathrm{flip}(\mathrm{mLSTM}(\mathrm{flip}(\bm{\varepsilon})))),
\end{align}
where $\bm{\varepsilon}$ is the input to the  time and frequency xLSTM blocks, and $\mathrm{mLSTM}(\cdot)$, $\mathrm{flip}(\cdot)$, $\oplus$, and $\mathrm{Conv1d}(\cdot)$ is the unidirectional mLSTM, the sequence flipping operation, concatenation, and the $1$-D transposed convolution, respectively.

Finally, the output of the TF-xLSTM blocks is decoded by both a magnitude mask decoder and wrapped phase decoder \cite{ai2023neural}. They predict the clean compressed magnitude mask $\bm{M}^c = (\bm{X}_m/\bm{Y}_m)^c\in\mathbb{R}^{T\times F}$ and the clean wrapped phase spectrum $\bm{X}_p\in\mathbb{R}^{T\times F}$, respectively. The enhanced magnitude spectrum $\bm{\hat{X}}_m\in\mathbb{R}^{T\times F}$ is obtained by computing:
\begin{align}
    \bm{\hat{X}}_m = ((\bm{Y}_m)^c \odot \bm{\hat{M}}^c)^{1/c},
\end{align}
where $\bm{\hat{M}}^c$ is the predicted clean compressed magnitude mask. 
The final enhanced waveform $\bm{\hat{x}}$ is reconstructed by performing an iSTFT on the enhanced magnitude spectrum $\bm{\hat{X}}_m$ and the enhanced wrapped phase spectrum $\bm{\hat{X}}_p$.

Similar to MP-SENet \cite{mp-senet} and SEMamba \cite{semamba}, we use a linear combination of loss functions which includes a PESQ-based GAN discriminator, along with time, magnitude, complex, and phase losses. We also employ the consistency loss function proposed in \cite{consistency_loss}, as this has been shown to improve performance \cite{semamba}.

\textbf{Feature encoder:} Following MP-SENet \cite{mp-senet}, the encoder consists of two convolution blocks each comprising a $2\mathrm{D}$ convolutional layer, an instance normalization, and  parametric rectified linear unit (PReLU) activation, sandwiching a dilated DenseNet \cite{densenet} with dilation sizes $1$, $2$, $4$, and $8$. The first convolutional block increases the input channels from $2$ to $C$, while the second convolutional block halves the frequency dimension from $F$ to $F'=F/2$, consequently reducing the computational complexity in the TF-xLSTM blocks. The dilated DenseNet extends the receptive field along the time axis, which facilitates long-range context aggregation over different resolutions.

\textbf{Magnitude mask and wrapped phase decoder:} Following MP-SENet \cite{mp-senet}, both the magnitude mask decoder and the wrapped phase decoder consist of a dilated DenseNet and a 2D transposed convolution. For the magnitude mask decoder, this is followed by a deconvolution block reducing the output channels from $C$ to 1. To estimate the magnitude mask, we employ a learnable sigmoid function with $\beta=2$ as in \cite{metricgan+}. In the wrapped phase decoder, the transposed convolution is followed by two parallel 2D convolutional layers outputting the pseudo-real and pseudo-imaginary part components. To predict the clean wrapped phase spectrum, we use the two-argument arctangent function (Arctan2) resulting in the enhanced wrapped phase spectrum $\bm{\hat{X}}_p$.

%% file: Text/03Experiments.tex
\section{Experiments}
\subsection{Dataset}
In this study, we perform experiments on the VoiceBank+Demand dataset, which consists of pairs of clean and noisy audio clips sampled at $\SI{48}{kHz}$. The clean audio samples come from the VoiceBank corpus \cite{voicebank}, which comprises $11,572$ audio clips from $28$ distinct speakers for training, and $824$ audio clips from $2$ distinct speakers for testing. The noisy audio clips are created by mixing the clean samples with noise from the DEMAND dataset \cite{demand} at four signal-to-noise ratios (SNRs) during training ([$0, 5, 10, 15$] dB) and testing ([$2.5, 7.5, 12.5, 17.5$] dB). Two speakers from the training set are left out as a validation set. 

\subsection{Implementation details}
Unless otherwise stated, experimental details and training configurations match those presented in MP-SENet \cite{mp-senet} and SEMamba \cite{semamba}. To reduce memory and computational resources, all models were trained on randomly cropped $2$-second audio clips. Additionally, all audio clips were downsampled to $\SI{16}{kHz}$, reducing computational complexity and ensuring compatibility with the wide-band PESQ metric \cite{pesq}. When performing STFTs we set the FFT order, Hann window size, and hop size  to $400$, $400$, and $100$, respectively. We train all models for $200$ epochs and select the checkpoint (saved every $1000$th step) with the best PESQ score on the validation data. We fix $C=64$ channels and $N=4$ stacks of TF-xLSTM blocks in our xLSTM-SENet model for direct comparison with SEMamba and MP-SENet. All models are trained with a batchsize $B=8$ on four NVIDIA L40S GPUs, and the four layer xLSTM-SENet model takes approximately $3$ days to train. This is the main limitation of xLSTM compared to Mamba and Transformers, which are roughly four times as fast to train \cite{xlstm}.

\subsection{Evaluation metrics}
We use the following commonly used evalutation metrics to assess SE performance: wide-band PESQ \cite{pesq} and short-time objective intelligibility (STOI) \cite{stoi}. To predict the signal distortion, background intrusiveness and overall speech quality, we use the composite measures CSIG, CBAK and COVL \cite{csigcbakcovl}. For all measures, a higher value is better. We train all models with $5$ different seeds and document the mean and standard deviation.

%% file: Text/04Results.tex
\section{Results and analysis}
\subsection{Comparison with existing methods}
We evaluate several architectural design choices for our xLSTM-SENet model while limiting the parameter count to that of SEMamba. We choose the best model based on validation performance. Our best model uses a bidirectional architecture and we have added biases to layer normalizations and projection layers as in \cite{visionlstm}. The expansion factor is set to $E_f=4$.
\begin{table}[H]
\centering
\caption{Results on the VoiceBank+Demand dataset. “-” denotes that the result is not provided in the original paper. $^*$ means the results are reproduced using the original provided code.}
\tabcolsep=0.1cm
\renewcommand*{\arraystretch}{1.1}
\begin{adjustbox}{width=1\columnwidth}
\begin{tabular}{lcccccc} 
\toprule
Model & \makecell{Params\\(M)}& PESQ & CSIG & CBAK & COVL & STOI\\
\midrule
Noisy & - & $1.97$ & $3.35$ & $2.44$ & $2.63$ & $0.91$\\
\midrule
MetricGAN+ \cite{metricgan+} & - & $3.15$ & $4.14$ & $3.16$ & $3.64$ & -\\
CMGAN \cite{cmgan} & $1.83$ & $3.41$ & $4.63$ & $3.94$ & $4.12$ & ${0.96}$\\
DPT-FSNet \cite{dpt}  & $0.88$ & $3.33$ & $4.58$ & $3.72$ & $4.00$ & ${0.96}$\\
Spiking-S4 \cite{spiking} & $0.53$ & $3.39$ & ${4.92}$ & $2.64$ & ${4.31}$ & -\\
TridentSE \cite{trident} & $3.03$ & $3.47$ & $4.70$ & $3.81$ & $4.10$ & ${0.96}$\\
MP-SENet \cite{mp-senet} & $2.05$ & $3.50$ & $4.73$ & ${3.95}$ & $4.22$ & ${0.96}$\\
SEMamba \cite{semamba} & $2.25$ & ${3.55}$ & $4.77$ & ${3.95}$ & $4.26$ & ${0.96}$\\
MP-SENet$^*$ & $2.05$ & $3.49\scriptstyle\pm 0.02$ & $4.72\scriptstyle\pm 0.02$ & $3.92\scriptstyle\pm 0.04$ & $4.22\scriptstyle\pm 0.02$ & ${0.96\scriptstyle\pm 0.00}$\\
SEMamba$^*$ & $2.25$ & $3.49\scriptstyle\pm 0.01$ & $4.75\scriptstyle\pm 0.01$ & $3.94\scriptstyle\pm 0.02$ & $4.24\scriptstyle\pm0.01$ & ${0.96\scriptstyle\pm0.00}$ \\
\midrule
xLSTM-SENet & $2.20$ & $3.48\scriptstyle\pm0.00$ & $4.74\scriptstyle\pm0.01$ & $3.93\scriptstyle\pm0.01$ & $4.22\scriptstyle\pm0.01$ & ${0.96\scriptstyle\pm0.00}$\\
\bottomrule
\end{tabular}
\end{adjustbox}
\label{tab:results}
\end{table}

\autoref{tab:results} shows that xLSTM-SENet matches the performance of the state-of-the-art SEMamba and MP-SENet models on the VoiceBank+Demand dataset, while outperforming other SE methods on most metrics. This demonstrates the effectiveness of xLSTM for SE.

\subsection{Ablation study}
To evaluate our model architecture design choices, we perform ablations on the expansion factor $E_f$ and on the biases in layer normalizations and projection layers. Additionally, we investigate the performance of a unidirectional architecture, by removing the transposed convolution, flipping and the second mLSTM block within each time and frequency xLSTM block as shown in \autoref{fig: SExLSTM}. Since our xLSTM-SENet is bidirectional and thus has 2 mLSTM blocks, we double the amount of layers $N$ for the unidirectional model. Finally, mLSTM adds exponential gating to improve LSTM, hence we evaluate its effect on speech enhancement performance by replacing it with sigmoid gating.
% mLSTM adds matrix memory and exponential gating to improve LSTM. To evaluate some of these improvements and our model architecture design choices, we perform ablations on the expansion factor $E_f$ and on the biases in layer normalizations and projection layers. Additionally, we evaluate the effect of exponential gating by replacing it with sigmoid gating. Finally, we investigate the performance of a unidirectional architecture, by removing the transposed convolution, flipping and the second mLSTM block within each time and frequency xLSTM block as shown in \autoref{fig: SExLSTM}, while doubling the amount of layers $N$ to match the parameter count of xLSTM-SENet. 
\begin{table}[H]
\centering
\caption{Ablation study on the VoiceBank+Demand dataset. Default settings for xLSTM-SENet: $E_f=4$, a bidirectional architecture, and biases in layer normalizations and projection layers.}
\tabcolsep=0.1cm
\renewcommand*{\arraystretch}{1.1}
\begin{adjustbox}{width=1\columnwidth}
\begin{tabular}{lcccccc} 
\toprule
Model & \makecell{Params\\(M)}& PESQ & CSIG & CBAK & COVL & STOI\\  
\midrule
Noisy & - & $1.97$ & $3.35$ & $2.44$ & $2.63$ & $0.91$\\
\midrule
xLSTM-SENet & $2.20$ & ${3.48\scriptstyle\pm0.00}$ & ${4.74\scriptstyle\pm0.01}$ & ${3.93\scriptstyle\pm0.01}$ & ${4.22\scriptstyle\pm0.01}$ & ${0.96\scriptstyle\pm0.00}$\\
$E_f=3$ & $1.96$& $3.46\scriptstyle\pm0.01$ & $4.72\scriptstyle\pm0.00$ & ${3.93\scriptstyle\pm0.02}$& ${4.21\scriptstyle\pm0.01}$ & ${0.96\scriptstyle\pm0.00}$\\
$E_f=2$ &$1.71$ &$3.45\scriptstyle\pm0.01$ & $4.71\scriptstyle\pm0.01$ & ${3.92\scriptstyle\pm0.01}$& $4.19\scriptstyle\pm0.01$ & ${0.96\scriptstyle\pm0.00}$\\
w/o Biases & $2.18$& $3.46\scriptstyle\pm0.00$ &  ${4.72\scriptstyle\pm0.01}$ &  ${3.91\scriptstyle\pm0.01}$ &  ${4.20\scriptstyle\pm0.02}$ & ${0.96\scriptstyle\pm0.00}$\\
Unidirectional & $2.14$& $3.26\scriptstyle\pm0.02$ & $4.57\scriptstyle\pm0.02$ & $3.79\scriptstyle\pm0.01$ & $4.00\scriptstyle\pm0.02$ & $0.95\scriptstyle\pm0.00$\\
w/o Exp. gating & $2.20$& $3.45\scriptstyle\pm0.02$ &${4.72\scriptstyle\pm0.02}$  & $3.90\scriptstyle\pm0.01$  & ${4.20\scriptstyle\pm0.03}$  & ${0.96\scriptstyle\pm0.00}$\\
\bottomrule
\end{tabular}
\end{adjustbox}
\label{tab:ablations}
\end{table}
\autoref{tab:ablations} shows that decreasing the expansion factor $E_f$ decreases performance. Moreover, as in Vision-LSTM \cite{visionlstm}, biases in layer normalizations and projection layers improve performance. We also find that a bidirectional architecture significantly outperforms a unidirectional architecture. Finally, we find that exponential gating improves performance for SE, which was not the case for learning self-supervised audio representations with xLSTMs \cite{audioxlstm}.

\subsection{Comparison with LSTM}
To compare the performance of xLSTM and LSTM for SE, we first replace the mLSTM layers in xLSTM-SENet with conventional LSTM layers (this models is referred to as: LSTM (layer)). \autoref{tab:lstmvsxlstm} shows that this results in a performance decrease even though LSTM (layer) is approximately \SI{11}{\percent} larger. Then, we replace the entire mLSTM block with LSTM (denoted as: LSTM (block)) and double the number of layers $N$ to roughly match the parameter count of xLSTM-SENet. \autoref{tab:lstmvsxlstm} shows that LSTM (block) matches xLSTM-SENet in performance, which in \autoref{tab:results} was shown to match the performance of state-of-the-art Mamba and Conformer-based systems.
\begin{table}[H]
\centering
\caption{Comparison of xLSTM with LSTM on the VoiceBank+Demand dataset.}
\tabcolsep=0.1cm
\renewcommand*{\arraystretch}{1.1}
\begin{adjustbox}{width=1\columnwidth}
\begin{tabular}{lcccccc} 
\toprule
Model & \makecell{Params\\(M)}& PESQ & CSIG & CBAK & COVL & STOI\\  
\midrule
Noisy & - & $1.97$ & $3.35$ & $2.44$ & $2.63$ & $0.91$\\
\midrule
xLSTM-SENet & $2.20$ & ${3.48\scriptstyle\pm0.00}$ & ${4.74\scriptstyle\pm0.01}$ & ${3.93\scriptstyle\pm0.01}$ & ${4.22\scriptstyle\pm0.01}$ & ${0.96\scriptstyle\pm0.00}$\\
LSTM (layer) & $2.44$& $3.44\scriptstyle\pm0.01$ & $4.69\scriptstyle\pm0.02$ & $3.90\scriptstyle\pm0.00$ & $4.17\scriptstyle\pm0.01$ & ${0.96\scriptstyle\pm0.00}$\\
LSTM (block) & $2.34$& ${3.49\scriptstyle\pm0.02}$ & ${4.76\scriptstyle\pm0.01}$ & ${3.95\scriptstyle\pm0.01}$ & ${4.24\scriptstyle\pm0.01}$ & ${0.96\scriptstyle\pm0.00}$\\
\bottomrule
\end{tabular}
\end{adjustbox}
\label{tab:lstmvsxlstm}
\end{table}

\subsection{Scaling experiments}
Smaller models are preferred in real-world SE applications, like hearing aids, due to reduced computational complexity, facilitating their use in such devices. Additionally, it is of interest to explore the performance achieved by increasing model sizes. Hence, we perform a comparative analysis of xLSTM, Mamba, Conformer and LSTM  across varying layer counts $N$. For LSTM, $N$ is doubled to roughly match the parameter counts of the xLSTM-, Mamba-, and Conformer-based models.

\begin{figure}[h]
    \centering \includegraphics[width=1\columnwidth]{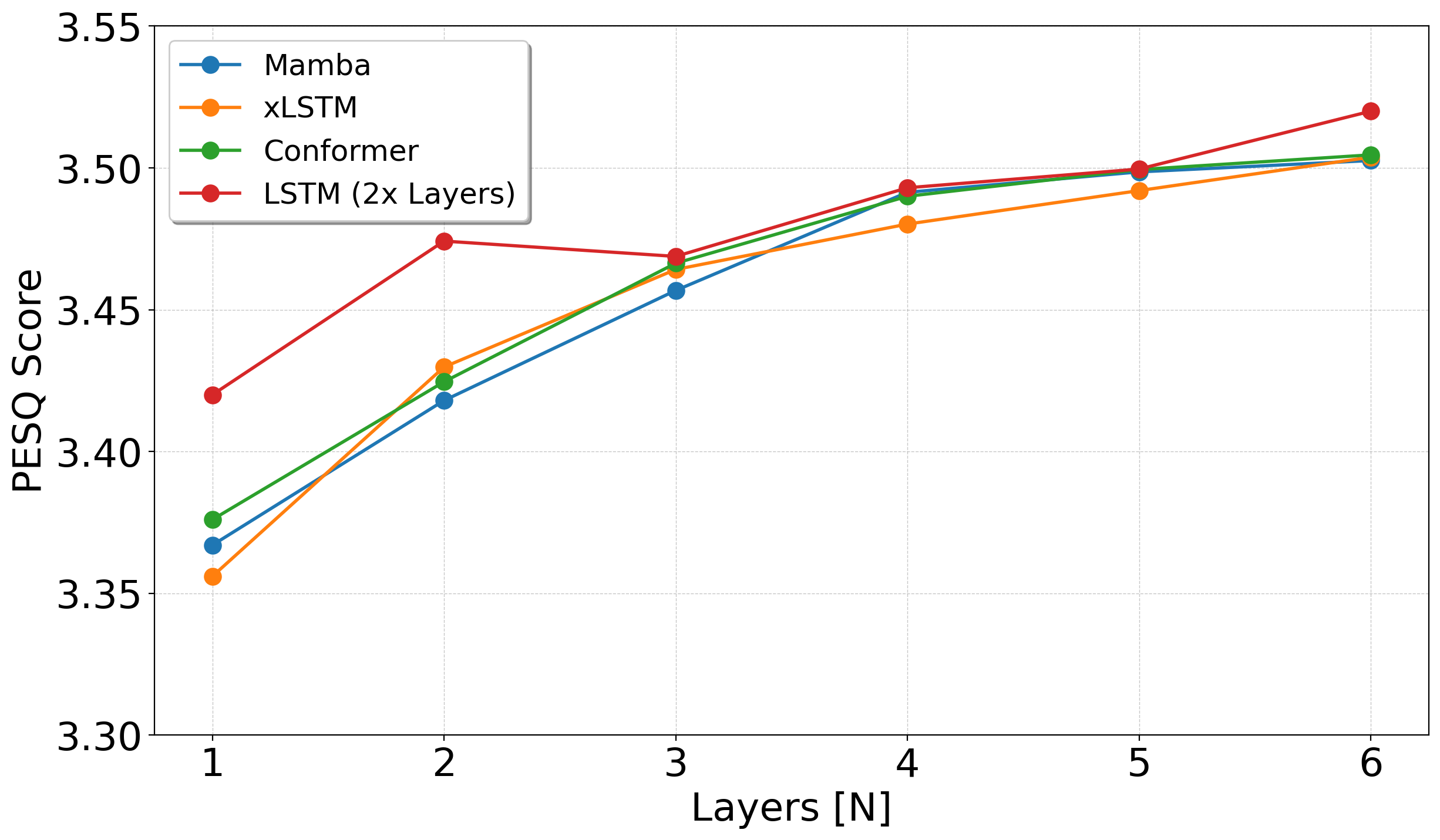}
    \caption{Scaling results on the VoiceBank+Demand dataset. The smallest ($N=1$) and largest ($N=6$) models are $\SI{1.37}{M}$ and $\SI{2.94}{M}$ parameters, respectively.}
    \label{fig:downscale}
\end{figure}
\autoref{fig:downscale} shows that xLSTM, Mamba, and Conformer-based models perform similarly when scaled down, with LSTM outperforming them for $N=1$ and $N=2$. When scaled up, all models achieve comparable performance.  For $N=1$, training-time is reduced by nearly \SI{70}{\percent} compared to $N=4$.

\subsection{xLSTM-SENet2}
To investigate the benefits of increased depth over width in xLSTM for SE, we propose xLSTM-SENet2, which is configured with an expansion factor $E_f=2$ and $N=8$ layers. This allows for a deeper architecture while maintaining a comparable parameter count to xLSTM-SENet. \autoref{tab:xlstmsenet2} shows that xLSTM-SENet2 outperforms state-of-the-art LSTM-, Mamba-, and Conformer-based models.
\begin{table}[H]
\centering
\caption{Speech enhancement performance of xLSTM-SENet2 on the VoiceBank+Demand dataset. $^*$ means the results are reproduced using the original provided code.}
\tabcolsep=0.1cm
\renewcommand*{\arraystretch}{1.1}
\begin{adjustbox}{width=1\columnwidth}
\begin{tabular}{lcccccc} 
\toprule
Model & \makecell{Params\\(M)}& PESQ & CSIG & CBAK & COVL & STOI\\  
\midrule
Noisy & - & $1.97$ & $3.35$ & $2.44$ & $2.63$ & $0.91$\\
\midrule
LSTM (block) & $2.34$& ${3.49\scriptstyle\pm0.02}$ & ${4.76\scriptstyle\pm0.01}$ & ${3.95\scriptstyle\pm0.01}$ & ${4.24\scriptstyle\pm0.01}$ & ${0.96\scriptstyle\pm0.00}$\\
MP-SENet$^*$ & $2.05$ & $3.49\scriptstyle\pm 0.02$ & $4.72\scriptstyle\pm 0.02$ & $3.92\scriptstyle\pm 0.04$ & $4.22\scriptstyle\pm 0.02$ & ${0.96\scriptstyle\pm 0.00}$\\
SEMamba$^*$ & $2.25$ & $3.49\scriptstyle\pm 0.01$ & $4.75\scriptstyle\pm 0.01$ & $3.94\scriptstyle\pm 0.02$ & $4.24\scriptstyle\pm0.01$ & ${0.96\scriptstyle\pm0.00}$ \\
\midrule
xLSTM-SENet2  & $2.27$ & $3.53\scriptstyle\pm 0.01$ & $4.78\scriptstyle\pm 0.01$ & $3.98\scriptstyle\pm 0.02$ & $4.27\scriptstyle\pm 0.01$ & $0.96\scriptstyle\pm 0.00$\\
\bottomrule
\end{tabular}
\end{adjustbox}
\label{tab:xlstmsenet2}
\end{table}

%% file: Text/05Conclusion.tex
\section{Conclusion}
This paper proposed xLSTM-SENet, an Extended Long Short-Term Memory-based model for speech enhancement. Experiments on the VoiceBank+Demand dataset show that xLSTM-SENet, and even LSTM-based models, rival existing state-of-the-art Mamba- and Conformer-based speech enhancement systems across several model sizes. We studied the importance of several architectural design choices, and demonstrated that the inclusion of exponential gating and bidirectionality is critical to the performance of the xLSTM-SENet model. Finally, empirical results show that our best xLSTM-based system, xLSTM-SENet2, outperforms state-of-the-art speech enhancement systems on the VoiceBank+Demand dataset.